\documentclass[fleqn,10pt]{wlscirep}
\title{Shortening time scale to reduce thermal effects in quantum transistors}

\usepackage[utf8]{inputenc}
\usepackage{amsmath}
\usepackage[normalem]{ulem}
\usepackage[caption = false]{subfig}
\usepackage{graphicx,epstopdf}
\usepackage{blindtext}
\usepackage{lipsum}
\usepackage{amsfonts}
\usepackage{bbm}
\usepackage{amssymb}
\usepackage{enumerate}
\usepackage{color}
\usepackage{latexsym}
\usepackage{times,txfonts}

\newcommand{\Fcal}{\mathcal{F}}

\newcommand{\ket}[1]{| #1 \rangle}

\newcommand{\bra}[1]{\langle #1 |}

\DeclareFontFamily{U}{mathc}{}
\DeclareFontShape{U}{mathc}{m}{it}%
{<->s*[1.03] mathc10}{}
\DeclareMathAlphabet{\mathscr}{U}{mathc}{m}{it}

\author[1,*]{M. A. de Ponte}
\author[2,$^{\dagger}$]{Alan C. Santos}
\affil[1]{Universidade Estadual Paulista (UNESP), Campus Experimental de Itapeva, 18409-010, Itapeva, S\~{a}o Paulo, Brazil}
\affil[2]{Instituto de F\'{i}sica, Universidade Federal Fluminense, Av. General Milton Tavares de Souza s/n, Gragoat\'{a}, 24210-346 Niter\'{o}i, Rio de Janeiro, Brazil}

\affil[*]{mickel.ponte@unesp.br}
\affil[$^{\dagger}$]{ac\_santos@id.uff.br}

\begin{abstract}
In this article, we present a quantum transistor model based on a network of coupled quantum oscillators destined to quantum information processing tasks in linear optics. To this end, we show in an analytical way how a set of $N$ quantum oscillators (data-bus) can be used as an optical quantum switch, in which the energy gap of the data bus oscillators plays the role of an adjustable ``potential barrier". This enables us to ``block or allow" the quantum information to flow from the source to the drain. In addition, we discuss how this device can be useful for implementing single qubit phase-shift quantum gates with high fidelity, so that it can be used as a useful tool. To conclude, during the study of the performance of our device when considering the interaction of this with a thermal reservoir, we highlight the important role played by the set of oscillators which constitute the data-bus in reducing the unwanted effects of the thermal reservoir. This is achieved by reducing the information exchange time (shortening time scale) between the desired oscillators. In particular, we have identified a non-trivial criterion in which the ideal size of the data-bus can be obtained so that it presents the best possible performance. We believe that our study can be perfectly adapted to a large number of thermal reservoir models.
\end{abstract}
\begin{document}

\flushbottom
\maketitle
%
%
\thispagestyle{empty}

\section*{Introduction}

Providing small devices that operate in quantum regime, maintaining high processing fidelity against the effects of decoherence, is of great importance in both quantum computing and information processing. Motivated by the
important role played by transistors in classical devices, we can imagine
how the quantum version of such electronic components could help us to achieve
even more significant advances in quantum information processing. In classical devices, a transistor can be used as a switch to block or
transfer classical information (encoded in terms of the intensity of
electric current, for example) from some source to a drain. On the other hand, differently from the classical transistor, because of the non-clone theorem \cite{Wootters:82}, a
quantum transistor can not copy arbitrary quantum information encoded in the
source. Therefore, by making an analogy with its classical counterpart, quantum transistors could be used to block or allow the flux of quantum information from a source to a drain. So that, to design
a quantum transistor, we must focus on the performance of the quantum switch,
trying to make it as efficient as possible. In this scenario, we need to study the transfer of quantum information between two
quantum systems (quantum bits - qubits). A greater motivation to design a quantum transistor is associated
with its applicability in quantum computation, as shown in the context of
adiabatic quantum computation~\cite{Bacon:17}, where fault-tolerant
universal quantum computation can be efficiently achieved if we can build an
``adiabatic quantum transistor". Besides that, the quantum transistors were
studied in the adiabatic quantum computing scenario~\cite{Williamson:15,Bacon:17}, spin chain~\cite{Marchukov:16,Loft:18},
ultra-cold atoms~\cite{Micheli:04,Vaishnav:08,Fuechsle:12}, and in the other
systems presented in many references~\cite{Gajdacz:14,Chang:07,Gardelis:99,Chen:13,Hwang:09,Bose:12}. As a contribution of this paper, we discuss how a bosonic quantum transistor could be designed by using a particular arrangement of coupled quantum harmonic oscillators, providing therefore a quantum device for blocking and/or transferring quantum information in linear optics.

In the literature there are similar works to what we aim to investigate here under the approach of Quantum State Transfer (QST)~\cite{Mickel:04-1,Mickel:04-2,Chen}. In the works~\cite{Mickel:04-1,Mickel:04-2}, the authors investigate some particular networks oscillators in the strong coupling regime, in which they verify that the transfer of some particular coherent states occur in a short time scale inversely proportional to the square root of the number of oscillators, as we verify here. Although the authors perceive that the effects of decoherence on this time scale are reduced, an analysis of the temperature effects is absent. In reference~\cite{Chen} the authors analyze the QST in a linear chain of $N$ constituents from the perspective of an adiabatic dynamics. In this work the authors also manage to inhibit the effects of decoherence to the situation in which the state of the reservoir is the vacuum. They show that the fidelity is so close to unity the smaller the ratio between the decay rate $\gamma$ and the coupling intensity between the constituents for a linear network with $N=39$ elements. In order to extend some of the existing works in the literature, we intend to investigate how the size of the data bus in a specific (and different) arrangement can be useful to inhibit the effects of decoherence in the presence of a thermal reservoir.

In this paper, we present a quantum transistor model that can be useful for quantum information processing in linear optics. To this end, we consider that two quantum oscillators (source and drain) are coupled to each other only indirectly through one or even a network composed of $N$ non-interacting quantum oscillators (data-bus) which play the role of the transistor gate (the quantum switch). Through the study carried out on this system, it is possible to demonstrate that the performance of quantum information blocking of our transistor is associated with the detuning between the resonant frequencies of the source and drain oscillators with the frequencies of the data-bus oscillators. By using the transistor we propose in this work, which is genuinely quantum, one show that when the gate is opened to transfer quantum information, we can adjust many parameters (frequencies, coupling strengths and the number of data-bus oscillators), or just a few, in order to implement simple qubit logical quantum gates associated to phase shift gates. In this sense, our study provides a model that makes logical quantum gates from quantum transistors, as an alternative to adiabatic quantum transistors~\cite{Bacon:17}. Finally, we study the performance against the decoherence of the quantum transistor model.

\section*{Results}

Any new quantum transistor proposal must be composed of three fundamental
parts: source, gate and drain. If we want to use it in quantum computing,
the most appropriate way would be to consider the source and drain as
two-level systems (a single qubit), whereas the gate represents a quantum
channel (consisting of one or more qubits) that indirectly connects the
source with the drain.

\begin{figure}
	\centering
	\includegraphics[scale=0.30]{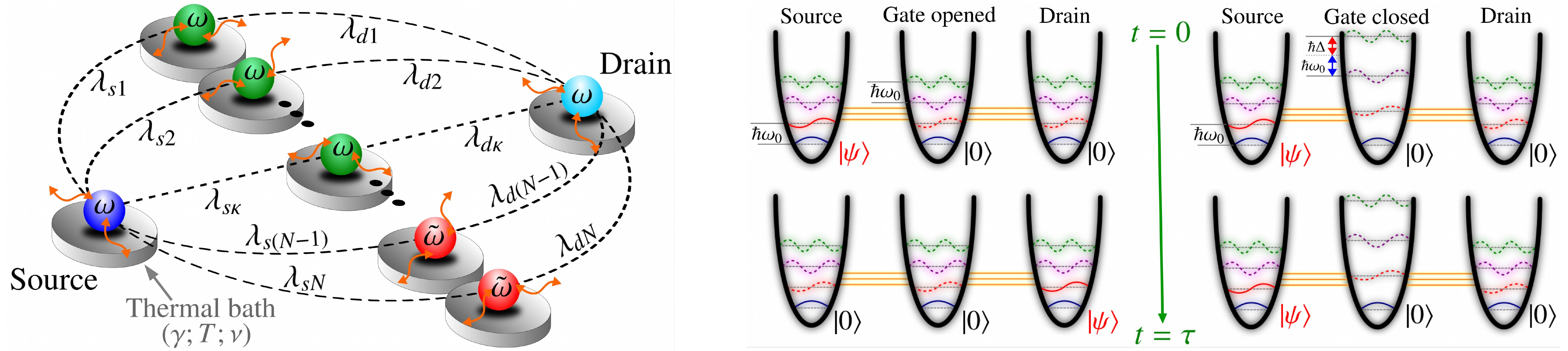}
	\caption{Left: Schematic representation of our quantum device, where the source and drain quantum oscillators are indirectly linked through a data bus. The data bus is constituted of $N$ quantum oscillators, in which $\kappa$ of them are at resonance with the source and drain (with frequency $\omega$), meanwhile $N - \kappa$ are far from resonance (with frequency $\tilde{\omega} \neq \omega$). Right: Scheme showing how the quantum gate (for the case of $\kappa = 1$) works, where the spacing between energy levels of the gate oscillator plays an important role for the performance of our transistor.}
	\label{Fig1}
\end{figure}

Let us consider hereafter that the subscript $\ell=s,d$, where $s$ ($d$) represents
the source (drain) oscillator, and the gate oscillators are labeled by $m$ and/or $n$. In general, considering a gate as a network of non-interacting quantum oscillators, instead of a single oscillator, the Hamiltonian that describes the dynamics of this system can be written as $H=H_{0}+V$, in which we defined 
\begin{eqnarray}
H_{0} =\hbar \sum \nolimits _{\ell }\omega _{\ell }a_{\ell }^{\dagger }a_{\ell
}+\hbar \sum \nolimits _{m}\omega _{m}a_{m}^{\dagger }a_{m},\quad\text{and}\quad V = \hbar \sum \nolimits _{m,\ell }\left( \lambda _{\ell m}a_{\ell }^{\dagger
}a_{m}+\lambda _{\ell m}^{\ast }a_{\ell }a_{m}^{\dagger }\right) \text{ , } \label{1}
\end{eqnarray}
such that $\omega$'s are the natural
frequencies of respective oscillators, whereas $\lambda$'s are the coupling strengths between two oscillators characterized by the pair of subscripted indices. The operators $a^{\dagger }$'s ($a$'s) represent the creation (annihilation) of a quanta in the respective oscillator. In Fig.~\ref{Fig1} we consider a transistor
composed of three parts, each part consisting of a quantum oscillator: the
source oscillator (the left), the drain oscillator (the right), and the gate
oscillator (in the middle) which is coupled to the first two.

\subsection*{The quantum transistor}

Since a quantum transistor must be able to control the quantum information
flow, we will consider, in our study, that the state $\left\vert \psi
_{s}\right\rangle $, of the source oscillator, has the information that can
be encoded in a quantum bit whereas the state of the drain oscillator and
that of the gate are in the vacuum state. In this way, the initial state of the
whole system can be written as a tensor product of the states of each
oscillator in the form $ \vert \Psi \left( 0\right)  \rangle
= \vert \psi _{s} \rangle \otimes \vert  \{ 0_{g} \}
 \rangle \otimes  \vert 0_{d} \rangle \equiv  \vert \psi
_{s}, \{ 0_{g} \} ,0_{d} \rangle $. It is known in the
literature that for a non-zero weak coupling between the resonant
oscillators that unknown information will flow from the source to the drain
in a transfer time $\tau _{\text{trans}}$ proportional to the coupling~\cite{Mickel:07-1,Mickel:07-2,Mickel:05}. The challenge of building a quantum
transistor can be achieved when we use a certain system parameter to allow
or block this transfer. If we could easily connect and disconnect the gate
couplings with the source and drain oscillators, this task would be
trivially executed in the situation where all oscillators are resonant.
However, in this work we are interested in nontrivial situations, in which
the couplings between oscillators are kept constant. To this end, we aim for
our quantum transistor to use our ability to increase or decrease the
frequency of one or more gate oscillators -- thus modifying the interval
between the energy levels of this oscillator -- so as to simulate the gate
in our device.

In order to discuss the behavior of our transistor for various situations of
interest, we will restrict ourselves to the parameter settings in Eq.~\eqref{1}
for the situation where the drain and source have same natural frequencies $\omega _{s}=\omega_{d}=\omega$ and the real coupling strengths between the oscillators are identical $\{\lambda_{sm}\}=\{\lambda_{gm}\}=\lambda$. As the main element of our system, the data bus configuration develops an important role in our transistor, as we shall see below. To illustrate the importance of this component, we will consider that $\kappa$ data-bus oscillators are
in resonance with the source and the drain, meanwhile the others $(N-\kappa )$ are out of resonance, such that we can write $\omega _{m} = \omega$, if $1\leq m\leq \kappa$, otherwise, $\omega _{m} = \tilde{\omega} = \omega +\Delta$, as shown in Fig.~\ref{Fig1}. Starting from the initial arbitrary state of the source oscillator $\left\vert \psi
_{s}\right\rangle =\alpha \left\vert 0_{s}\right\rangle +\beta e^{i\theta
}\left\vert 1_{s}\right\rangle $ and assuming that the time evolution can be performed by the operator $U(t)=e^{-iHt/\hbar }$, then we can conclude that the time evolution of the initial state, using the fact that $H \vert 0_{s}, \{ 0_{g} \} ,0_{d} \rangle=0 $, will be given by
\begin{equation}
\left\vert \Psi \left( t\right) \right\rangle =  \alpha \vert
0_{s}, \{ 0_{g} \} ,0_{d} \rangle +\beta e^{i\theta }U\left( t\right) \vert
1_{s}, \{ 0_{g} \} ,0_{d} \rangle \mathrm{ . }  \label{2}
\end{equation}

From Eq.~\eqref{2} it is easy to show that the probabilities $p_{s}(t)$ and $%
p_{d}(t)$ of finding the original information in the qubit source and drain
are, respectively, given by%
\begin{eqnarray}
p_{s}(t) = \left\vert \langle \Psi \left( 0\right) \vert
\Psi \left( t\right) \rangle \right\vert ^{2} 
 = \left\vert \alpha
^{2}+\beta ^{2} \langle 1_{s}, \{ 0_{g} \} ,0_{d} \vert
U\left( t\right)  \vert 1_{s}, \{ 0_{g} \} ,0_{d} \rangle
\right\vert ^{2}\mathrm{ , }  \label{BlockPro} \\
p_{d}(t) = \left\vert \langle \Phi \left(0 \right) \vert
\Psi \left( t\right) \rangle \right\vert ^{2} 
= \left\vert \alpha
^{2}+\beta ^{2} \langle 0_{s}, \{ 0_{g} \} ,1_{d} \vert
U\left( t\right) \vert 1_{s}, \{ 0_{g} \} ,0_{d} \rangle
\right\vert ^{2}\mathrm{ , }  \label{TransferPro}
\end{eqnarray}%
where we define the state $ \vert \Phi \left(0 \right)  \rangle
= \vert 0_{s}, \{ 0_{g} \} ,\psi _{d} \rangle = \vert
0_{s}, \{ 0_{g} \}  \rangle \otimes ( \alpha \vert
0_{d} \rangle +\beta e^{i\theta } \vert 1_{d} \rangle )$.

\subsubsection*{Blocking and transferring quantum information}

From Eq.~\eqref{BlockPro} it is possible to show that $p_{s}(t)=1$ and $%
p_{d}(t)=|\alpha |^{4}$ if, and only if, the matrix element defined by $%
u_{+}(t)= \langle 1_{s}, \{ 0_{g} \} ,0_{d} \vert U\left(
t\right) \vert 1_{s}, \{ 0_{g} \} ,0_{d} \rangle =1$ and, simultaneously, $%
u_{-}(t)= \langle 0_{s}, \{ 0_{g} \} ,1_{d} \vert U\left(
t\right)  \vert 1_{s}, \{ 0_{g} \} ,0_{d} \rangle =0$.
For the parameter regime we are considering, it is possible to perform the analytical calculation to determine both eigenvalues and
eigenvectors in order to obtain the expressions (see methods section for more details) $u_{\pm }(t)=[\Lambda \left( t\right) \pm 1]e^{-i\omega t}/2$, where we defined
\begin{equation}
\Lambda (t)=\mathcal{A}_{0}e^{-iR_{0}t/3}+\mathcal{A}_{+}e^{-iR_{+}t/3}+%
\mathcal{A}_{-}e^{-iR_{-}t/3},
\end{equation}
the amplitudes (for $\{j=0,+,-\}$)
\begin{equation*}
\mathcal{A}_{j}= \left[ 1+2\kappa \left( \frac{3\lambda }{R_{j}}\right)
	^{2}+2(N-\kappa )\left( \frac{3\lambda }{R_{j}-\Delta }\right) ^{2} \right]^{-1} \text{ , }
\end{equation*}%
the coefficients $R_{0} = \Delta +2\sqrt{\Delta ^{2}+6N\lambda ^{2}}\cos \theta$ and $R_{\pm } = \Delta -\sqrt{\Delta ^{2}+6N\lambda ^{2}}\left( \cos \theta \pm 
\sqrt{3}\sin \theta \right)$, and the angle $\theta$ by
\begin{equation*}
\theta =\frac{1}{3}\arctan \left( \sqrt{\frac{\left( \Delta ^{2}+6N\lambda
		^{2}\right) ^{3}}{\Delta ^{2}\left[ \Delta ^{2}+9\left( N-3\kappa \right)
		\lambda ^{2}\right] ^{2}}-1}\right) \text{.}
\end{equation*}
In this way, if we consider the case where $\lambda /\Delta \ll \left( 3%
\sqrt{N}\right) ^{-1}$, disregarding quadratic or superior terms, we can
approximate $\sin \theta \approx \theta \approx \sqrt{6\kappa }\lambda
/\Delta $ and $\cos \theta \approx 1$, and, consequently, the frequencies
can be approximate to the values $R_{0}\approx 3\Delta $, $R_{\pm }\approx
\mp 3\sqrt{2\kappa }\lambda $ and the amplitudes to $\mathcal{A}_{0}\approx
0 $ and $\mathcal{A}_{\pm }\approx 1/4$. With these values in hand, one can
easily determine the quantities 
\begin{eqnarray}
u_{+}(t)  = e^{-i\omega t}\cos ^{2}\left( \sqrt{\frac{\kappa }{2}}\lambda
t\right),\quad\text{and}\quad
u_{-}(t) = -e^{-i\omega t}\sin ^{2}\left( \sqrt{\frac{\kappa }{2}}\lambda
t\right) \text{.}  \label{RT}
\end{eqnarray}

Note from Eq.~\eqref{RT} that when none of the data bus oscillators is resonant with the source and drain (i.e., $\kappa =0$), we get $u_{+}(t)=e^{-i\omega t}$ and $u_{-}(t)=0$. This result shows that in the
regime where $\lambda /\Delta \ll ( 3\sqrt{N} ) ^{-1}$, the system
dynamics becomes the same as a single isolated oscillator evolving over
time. The role of this oscillating phase in Eq.~\eqref{RT} can be better understood when we write the evolution of the state $\left\vert \Psi \left(
0\right) \right\rangle $ in the regime $\lambda /\Delta \ll ( 3\sqrt{N}) ^{-1}$ with $\kappa =0$, that turns to be
\begin{eqnarray}
\left\vert \Psi \left( t\right)
\right\rangle \approx \left( \alpha \left\vert 0_{s}\right\rangle +\beta
e^{i\left( \theta -\omega t\right) }\left\vert 1_{s}\right\rangle \right)
\otimes \vert \{ 0_{g} \} ,0_{d} \rangle \text{ . }
\end{eqnarray}
Note that the information is maintained at the source, despite the appearance of a
time-dependent local phase. As the value of this phase varies, we may have $%
 ( 1-2\beta ^{2} ) ^{2}\leq p_{s}(t)\leq 1$, which does not mean
that the information flows from the source to the drain (or be partially
destroyed), once the drain state
remains in the vacuum, as we can see from $u_{-}(t)=0$. Moreover, it is important to highlight the fact that $p_{d}(t) = |\alpha |^{4}$ and not zero. This value is not null because the information contained in the state $\left\vert \Psi \left(
0\right) \right\rangle $ has a component $\left\vert 0\right\rangle $, whose
probability amplitude is $\alpha =\left\langle 0\right. \left\vert \Psi
\left( 0\right) \right\rangle $, regardless of whether or not there is a
dynamic between the source and drain oscillators.

Despite this, the performance of our model is not affected by this
``unwanted" phase. In addition, if we let the system evolve indefinitely,
whenever time $t$ is a positive integer, $n$, multiple of the recurrence
time $\tau _{R}=2\pi /\omega $, we get exactly the input state $\left\vert
\psi _{s}\right\rangle $ encoded in the source qubit. Further on, we will
see that this ``unwanted" phase becomes indispensable if we are to use our
device to implement quantum gates. Therefore, using the scheme in Fig. \ref{Fig1}, it is possible to use
the $\Delta $-dissonance to block the quantum information indefinitely.
Since the blocking situation is associated with our ability to adjust $%
\lambda /\Delta \ll 1/ (3\sqrt{N})$, our model can be
efficiently implemented using quantum dot-cavity systems~\cite{Hennessy:07,Li:02}, coupled-cavity array~\cite{Zhang:17,Irish:08,Xiao:08},
bosonic lattice systems~\cite{Eisert:15,Kaufman:16} or cold atoms~\cite{Tomita:17}, for example.

To analyze the quantum information transfer it is desirable to imagine that the gate configuration of the
transistor (open or closed) should be controlled by the adjustment of a
single physical parameter, otherwise we may have some technical difficulties
in handling a set of parameters. In this way, as we use the dissonance $%
\Delta $ to close the gate, we need to show how this same parameter could be
used to open it. In other words, from Eq. (\ref{RT}) we must discuss how a
new adjustment of $\Delta $ allows us to obtain $p_{d}(t)=1$. If we want
that the state $\left\vert \Psi \left( 0\right) \right\rangle $ will be
transferred to the drain oscillator, two adjustments must be made
simultaneously: $e^{-i\omega t}=-1$ and $\sin ^{2} ( \sqrt{\kappa /2}%
\lambda t ) =1$. These adjustments imply that $\omega t$ and $\sqrt{\kappa/2}\lambda t$ must be an odd number multiple of $\pi $  and $\pi /2$, respectively. That is, we need to have $t=\left( 2j+1\right) \pi /\omega $ and $t=\left( 2j^{\prime
}+1\right) \pi / ( \lambda \sqrt{2\kappa } ) $, respectively, with $j$ and $j^{\prime
}$ integers. To find the exact value of the transfer time $\tau _{\text{trans}}$, we must find the integers $j$ and $j^{\prime }$ that satisfy the equality
\begin{equation}
\tau _{\text{trans}}=\frac{2j+1}{\omega }\pi =\frac{2j^{\prime }+1}{\lambda 
	\sqrt{2\kappa }}\pi \text{ . }  \label{I}
\end{equation}%

Note that this equality can only be satisfied when $\lambda \sqrt{2\kappa }%
/\omega $ is the ratio between two odd numbers $C_{1}/C_{2}$, which can always be achieved, regardless of the values of $\lambda $ and $\omega$, with the convenient adjust of $\kappa $. Once we have made this adjustment, the transfer time becomes
\begin{equation*}
\tau _{\text{trans}}=\frac{2j+1}{\omega }\pi \qquad \text{or}\qquad \tau _{\text{trans}}= \frac{%
	2j^{\prime }+1}{\lambda \sqrt{2\kappa }}\pi \text{ , } 
\end{equation*}%
where we must choose the smallest values of $j$ or $j^{\prime }$ for which $
C_{1}\left( 2j+1\right) =C_{2}\left( 2j^{\prime }+1\right) $.

To better understand this adjustment of $\kappa$, let us consider a numerical example. For the case where $\omega =10^{10}$ Hz and $
\lambda =10^{4}$ Hz, we can adjust $\kappa =2^{11}=2048$ in order to eliminate the powers of two from the decomposition in prime numbers of $ \lambda $ and $ \omega $ and so $\lambda \sqrt{2\kappa }/\omega $ becomes the ratio between two odd numbers. For this choice, we obtain $C_{1}=1$ and $
C_{2}=5^{6}$ and consequently we have $j^{\prime }=0$ and $j=\left(
10^{6}-1\right) /2=7812$. With these values, the transfer time will be
multiple integers of $\tau _{\text{trans}}=\pi /64\lambda $. Another important point to highlight is the following: since $e^{-i\omega t}$ is a function that oscillates very quickly when compared to $\sin^{2} \left(\sqrt{\kappa/2}\lambda t \right)$, we observe the existence of secondary peaks approaching the unit at time $t_{\text{ex}}=\pi/(\lambda\sqrt{2\kappa})$, such that $\sin^{2} \left(\sqrt{\kappa/2}\lambda t_{\text{ex}} \right)=1$. This characteristic time of the system, which represents the information exchange time between the source and drain oscillator, is inversely proportional to $\lambda\sqrt{2\kappa}$, so that the larger $\kappa$, the shorter $t_{\text{ex}}$. (In the literature the characteristic time of the system is sometimes called a short time scale, i.e., it is the minimum time required for there to be a significant change in the state due to the dynamics of the system. When we focus on the interaction picture, this time is proportional to the inverse of the coupling and when we look at a network system interacting, this time also becomes to be proportional to the square root of the number of oscillators N. \cite{Mickel:04-1,Mickel:04-2}) This reduction of time will be the key point to the study of the performance of our device against the effects of the thermal reservoirs, as we will see later.

\subsubsection*{Detuning control with atom-field interaction}

In order to obtain a optimum control of our transistor it is necessary that we are
able to adjust the detuning parameter $\Delta $ between the data bus
oscillators frequencies with the source and drain oscillators. This
control can be accomplished through a dispersive interaction\cite{Rev06} between an atom and the field inside the cavity, for example. 

To illustrate this procedure, we consider the Hamiltonian $%
H_{\text{disp}}=H_{\text{field}}+H_{\text{atom}}+H_{\text{atom-field}}$, where $H_{\text{field}}=\hbar \omega
_{0}a_{\ell }^{\dagger }a_{\ell }$ is the free Hamiltonian of a single
interacting mode $\ell $, $H_{\text{atom}}=\hbar \nu \sigma _{z}$ is the
Hamiltonian of a two-level atom and $H_{\text{atom-field}}=\hbar \chi a_{\ell
}^{\dagger }a_{\ell }\ \sigma _{3}$ is the Hamiltonian of the dispersive
interaction between the atom and the field, where $\sigma _{z}= \vert
e \rangle  \langle e \vert - \vert g \rangle
 \langle g \vert $ and $\sigma _{3}=\left\vert i\right\rangle
\left\langle i\right\vert -\left\vert e\right\rangle \left\langle
e\right\vert $. $\left\vert e\right\rangle $ ($\left\vert i\right\rangle $)
denotes the excited (virtual intermediate) state of the atom. The constant $%
\chi =g^{2}/\delta $ is given in terms of the atom-field coupling intensity, 
$g$, and the detuning $\delta =\omega _{0}-\nu $ between the field and atom
frequencies. It is important to remember that the validity of this dispersive Hamiltonian is confined in situations where $g^2 n\ll \delta^2+\gamma^2$, where $n$ is the mean number of photons in the field and $\gamma$ is the spontaneous emission rate. The time evolution of an atom-field state, according to the
Hamiltonian $H_{\text{disp}}$, will be given by the operator $U_{\text{disp}}\left(
t\right) =e^{-iH_{\text{disp}}t/\hbar }$. Since the commutator $ [
a_{\ell }^{\dagger }a_{\ell }\left( \omega _{0}+\chi \sigma _{3}\right)
,\sigma _{z} ] =0$, we can decompose the time evolution between $%
H_{\text{field}}+H_{\text{atom-field}}$ and $H_{\text{atom}}$, so that for an initial state of
the atom-field system given by $\left\vert \psi _{\text{atom-field}}\left( 0\right)
\right\rangle =\left( a\left\vert 0\right\rangle +b\left\vert 1\right\rangle
\right) \otimes \left\vert e\right\rangle $, we obtain the following state
evolved in time 
\begin{eqnarray}
\left\vert \psi _{\text{atom-field}}\left( t\right) \right\rangle = e^{-\frac{i}{\hbar} H_{\text{field}} t }e^{-\frac{i}{\hbar}( H_{\text{field}}+H_{\text{atom-field}})
	t }\left\vert e\right\rangle \otimes \left( a\left\vert 0\right\rangle
+b\left\vert 1\right\rangle \right)  = e^{-i\left( \omega _{0}-\chi \right) a_{\ell }^{\dagger }a_{\ell }}\left(
a\left\vert 0\right\rangle +b\left\vert 1\right\rangle \right) \otimes
\left( e^{-\frac{i}{\hbar}H_{\text{field}}t }\left\vert e\right\rangle \right) \text{ . }  \label{EQI}
\end{eqnarray}%

Thus, one can see from (\ref{EQI}) that the field state is factorized and can therefore
be discarded at the end of the process. Moreover, we conclude that the
temporal evolution of the field can be determined by the effective
Hamiltonian%
\begin{equation*}
\tilde{H}_{\text{field}}=\hbar \left( \omega _{0}-\chi \right) a_{\ell }^{\dagger
}a_{\ell }\text{ , }
\end{equation*}%
demonstrating that the field behaves effectively with a shift in the energy $%
\omega _{0}\rightarrow \omega _{0}-\chi $, when it interacts dispersively
with the atom. The same result can be verified when we take into account the
interaction between the quantum oscillators.

\subsection*{Application to quantum computation}

In general, the conditions previously discussed (for transferring and blocking
quantum information) lead us to think about what happens if we ignore
them. As we will show in this section, by violating the condition $\omega
t = \left( 2j+1\right) \pi $ (for $j=0,1,2,\ldots $), but
maintaining the condition $\lambda t = \left(
2j^{\prime}+1\right) \pi /\sqrt{2\kappa }$ (for $j^{\prime}=0,1,2,\ldots $%
), we can implement quantum phase-shift gates. In particular, we are
interested in a situation where we simultaneously transfer the information
and apply a quantum gate, such that we will define $\kappa >0$ hereafter.

In order to demonstrate how the quantum transistor we propose in this paper
allows us to implement a particular set of quantum gates, let us consider the system input state as 
$\vert \psi _{s}, \{ 0_{g} \} ,0_{d} \rangle $. We know that at
time $t_{\text{ex}}=\pi /(\lambda\sqrt{2\kappa })$, the system output state is given by
\begin{equation}
\vert \Psi \left( t_{\text{ex}}\right) \rangle = \vert 0_{s}, \{
0_{g}\}  \rangle \otimes \left( \alpha  \vert
0_{d} \rangle -\beta e^{i [ \theta -\omega \pi / ( \lambda 
	\sqrt{2\kappa } ) ] } \vert 1_{d} \rangle \right) \text{ . } \label{gate} 
\end{equation}
From Eq.(\ref{gate}), it can be seen that the output state is identical to the input state, except for a local phase that must be applied to the state $\vert 1 \rangle$. This result resembles that obtained by the phase shift gates $R\left( \phi \right) $, which are single-qubit gates that can be combined with other one- and/or two-qubit gates to provide a set of universal quantum gates \cite{Nielsen:Book,Barenco:95}. In general, given
any input state $\vert \psi _{\text{inp}} \rangle =\alpha \left\vert
0\right\rangle +\beta e^{i\theta }\left\vert 1\right\rangle $, after the
unitary operation $R\left( \phi \right) $, the output state becomes $
\left\vert \psi _{\text{out}}\left( \phi \right) \right\rangle =R\left( \phi
\right) \vert \psi _{\text{inp}} \rangle =\alpha \left\vert
0\right\rangle +\beta e^{i\left( \theta +\phi \right) }\left\vert
1\right\rangle $, for any real $\phi$. Thus, by considering the result in Eq.~\eqref{gate} and the properties of $R\left( \phi \right) $, one can see that the drain output state becomes $\left\vert \psi
_{\text{out}}\left( \phi \right) \right\rangle =R\left( \phi \right) \vert
\psi _{\text{inp}} \rangle $, when we performed the adjustment for $\omega/(\lambda\sqrt{2\kappa})$ given by
\begin{equation}
\frac{\omega}{\lambda\sqrt{2\kappa}}=\ell -\frac{\phi }{\pi }>0,  \label{3}
\end{equation}%
where $\ell$ must be odd.

For any given $\phi $, the expression (\ref{3}) above shows us how we should
make the adjustment in $\omega $, if $\lambda $ and $\kappa $ are fixed. As a first important remark of the data bus role in our device, in case we have a physical system in which $\omega$ and $\lambda$ are fixed, for example in cavity QED, we can choose the best fit of the integer $\kappa $ in order to implement the gate. It is important to note that the equality in (\ref{3}) can be obtained without any restriction with respect to the weak ($\lambda N\ll \omega $) or strong ($\lambda N\approx \omega $) coupling regime due to the presence of the term $\ell $, which can be an odd large or small number.

In conclusion, with the adjustment made in Eq.~\eqref{3}, the final state becomes
\begin{equation*}
\left\vert \Psi \left( t_{\text{ex}} \right) \right\rangle =\vert 0_{s},\{
0_{g}\} \rangle \otimes \left[ \alpha \left\vert
0_{d}\right\rangle + \beta e^{i(\theta +\phi )}\left\vert 1_{d}\right\rangle %
\right] = \vert 0_{s},\{
0_{g}\} \rangle \otimes \left[R\left( \phi \right)  \vert
\psi _{\text{inp}} \rangle \right]\text{ . }
\end{equation*}

Therefore, that the control can be done by simultaneous adjustment of $\omega$, $\kappa$ and, whenever available, the parameter $\lambda$. That shows that the quantum transistor proposed here can be used to make logic quantum gates, in the same way that classical transistors can implement logic classical gates.

\subsection*{Performance against decoherence}

In order to study the performance of our quantum transistor model against
the decoherence effects, we will consider that the system is coupled to
dissipative reservoirs according to a Lindblad equation \cite{Lindblad:76}. As shown in Fig.~\ref{Fig1}, in a \textit{quasi} realistic scenario, each oscillator of our system is evolving under action of individual thermal baths, where each one is at temperature $T$ and it is constituted by a infinite set of oscillators whose the average value of the frequency is around to $\nu$. In this case, the dynamics of the system can be written~\cite{Mickel:04-1,Mickel:04-2,Mickel:05,Mickel:07-1} as
\begin{equation}
\dot{\rho}(t) =\frac{1}{i\hbar }[H,\rho (t)]+\mathcal{L}_{\mathrm{se}%
}[\rho (t)]+\mathcal{L}_{\mathrm{th}}[\rho (t)]\mathrm{\ ,}  \label{LindEq}
\end{equation}%
where the operators are defined by
\begin{eqnarray}
\mathcal{L}_{\mathrm{se}}[\bullet ]  = \sum_{k}\frac{\gamma _{k}}{2}\left[
2a_{k}\bullet a_{k}^{\dagger }-\{a_{k}^{\dagger }a_{k},\bullet \}\right] 
\mathrm{ , } \quad 
\mathcal{L}_{\mathrm{th}}[\bullet ] =\sum_{k}\frac{\gamma _{k}\bar{n}_{k}}{%
	2}\left[ 2a_{k}\bullet a_{k}^{\dagger }-\{a_{k}^{\dagger }a_{k},\bullet \}+%
\mathrm{h.c.}\right] \mathrm{.}  \label{th}
\end{eqnarray}%
The operator $\mathcal{L}_{\mathrm{se}}[\bullet ]$ is associated with spontaneous emission effects while the operator $\mathcal{L}_{\mathrm{th}}[\bullet ]$ takes into account the dispersion process associated with a thermal reservoir at temperature $T\neq 0$. Here $\gamma _{k}$ is the emission rate of the $k$-th oscillator ($k=s,d,1,2,3,\ldots $), $\bar{n}_{k}$ is the average number of thermal photons in the $k$-th reservoir as calculated from the Planck distribution with $\bar{n}=1/(e^{h\nu /k_{\mathrm{B}}T}-1)$, and $k_{\mathrm{B}}$ is the Boltzmann constant. In particular, it is worth mentioning that non-unitary effects on quantum transistor has been addressed in superconducting quantum transistor models~\cite{Loft:18}, where the authors considered the transfer performance of the device against dephasing noise. Thus, the effects of thermal baths on such models is yet a open question.

Since our main interest is in the final state of the drain oscillator, where we will find the quantum information derived from the logic gate result, the state's fidelity will be computed through $\Fcal=\bra{\psi_{\text{out}}}\rho(t_{\text{ex}})\ket{\psi_{\text{out}}}$, where $\rho(t_{\text{ex}})$ is the whole density matrix of the system which comes from the solution of Eq.(\ref{LindEq}) while $\ket{\psi_{\text{out}}} = \ket{0_{\text{s}},\{0_{\text{g}}\}} \otimes R\left( \phi \right)\ket{\psi}$ is the ideal output state resulting of the logical operation $R\left( \phi \right) $ on a arbitrary input single qubit state $\ket{\psi} = \alpha \left\vert 0\right\rangle +\beta e^{i\theta
}\left\vert 1\right\rangle$ encoded in drain oscillator. 

For the regime of parameters we are considering in this work ($\omega_s=\omega_d=\omega$ and $\{\lambda_{sm}\}=\{\lambda_{dm}\}=\lambda$) it is possible to obtain an analytical solution of master equation Eq.~\eqref{LindEq} for the initial state in which the information is encoded in the source oscillator. If furthermore we also consider that all of data-bus oscillators are at resonance with the source and drain oscillators, that is, $\kappa = N$, the fidelity of finding the desired output state from the unit operation $R(\phi)$, encoded in the drain oscillators, is written as (see Method Section)
\begin{eqnarray}
\Fcal = \left[ 1 + \bar{n} \left(1- e^{ - \frac{\pi\gamma}{\lambda \sqrt{2\kappa}}}\right)  \right]^{-(3+\kappa)} \left[ \bar{n} + \alpha^2 +2\alpha^2\left(1-\alpha^2\right) e^{ - \frac{\pi\gamma}{2\lambda \sqrt{2\kappa}}} + \frac{2 (1-\alpha^2)^2}{(1+\bar{n})e^{ \frac{\pi\gamma}{\lambda \sqrt{2\kappa}}}-\bar{n}} + e^{-\frac{\pi\gamma}{\lambda \sqrt{2\kappa}}} \left( \alpha^2 - 1 - \bar{n} \right)\right] \text{ , } \label{Fifel}
\end{eqnarray}
where we already used the normalization condition $\alpha^2+\beta^2 = 1$ and we set $\omega$ as provided by Eq.~\eqref{3}. As expected, $\Fcal$ depends on the initial state and so that it is convenient to define an average value $\bar{\Fcal} = \langle \Fcal \rangle_{\psi} $ over all initial state $\ket{\psi}$. Therefore, we have
\begin{eqnarray}
\bar{\Fcal} = \left[ 1 + \bar{n} \left(1- e^{ - \frac{\pi\gamma}{\lambda \sqrt{2\kappa}}}\right)  \right]^{-(3+\kappa)} \left[ \bar{n} +\frac{1}{3} + \frac{4}{15} e^{ - \frac{\pi\gamma}{2\lambda \sqrt{2\kappa}}} - \left(\frac{2}{3}+\bar{n}\right)e^{ - \frac{\pi\gamma}{\lambda \sqrt{2\kappa}}} + \frac{16}{15}\frac{1}{ (1+\bar{n})e^{\frac{\pi\gamma}{\lambda \sqrt{2\kappa}}} -\bar{n} }\right] \text{ . } \label{Fifelbar}
\end{eqnarray}

\begin{figure}[t!]
	\centering
	\includegraphics[scale=0.6]{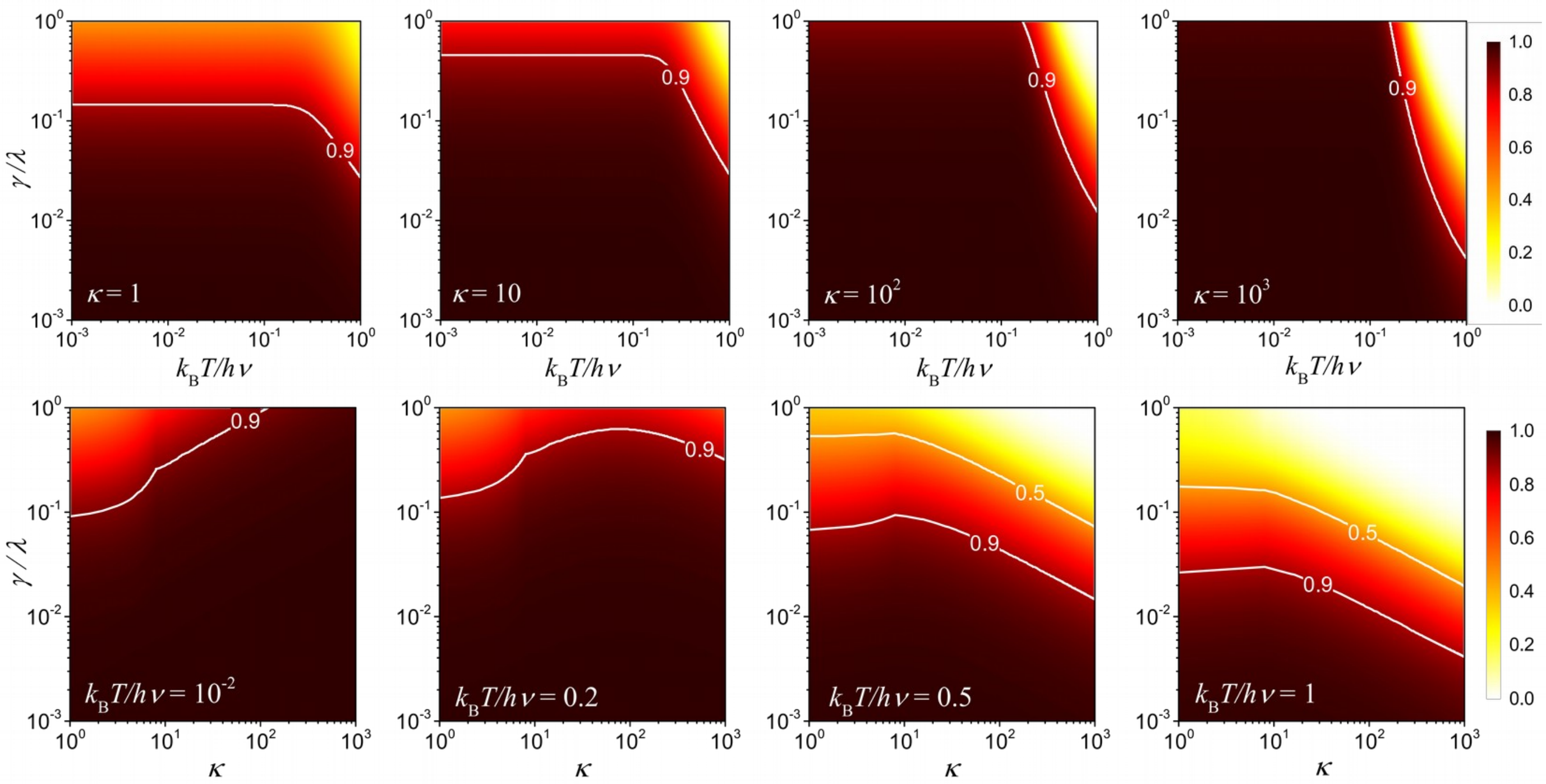}
	\caption{Density plot for $\bar{\Fcal}$ according to two sets of dimensionless parameters: (top) as a function of the ratio between the emission rate $\gamma$ with the coupling strength $\lambda$ and the temperature of the thermal baths through the ratio $k_{\text{B}}T/h\nu$, and (bottom) as a function of $\gamma/\lambda$ and the number $\kappa$ of resonant data-bus oscillators for different values of $k_{\text{B}}T/h\nu$. Note that the optimality criteria of $\bar{\Fcal}$, concerning the parameter $\kappa$, becomes evident with the highlight for the line that divides the densities regions larger and smaller than $0.9$ in the graphs with $k_{\text{B}}T/h\nu = 2\cdot10^{-1}$, $k_{\text{B}}T/h\nu = 5\cdot10^{-1}$ and $k_{\text{B}}T/h\nu = 1\cdot10^{0}$.}
	\label{Fig2}
\end{figure}

The non-trivial form of $\bar{\Fcal}$ with respect to the physical parameters involved does not allow us to find optimal strategies to analyze the effects of noisy environment through an analytical approach. For this reason, we consider a numerical study of the behavior of $\bar{\Fcal}$ as given in the density graph shown in Fig.~\ref{Fig2}. Firstly, in Fig.~\ref{Fig2}~(top) we present the behavior of $\bar{\Fcal}$ as a function of the dimensionless parameters $\gamma/\lambda$ and $k_{\text{B}}T/h\nu$, which are associated to the reservoir parameters $\gamma$ and $T$, for different values of the data-bus size $\kappa$. The range of values considered here for the quantity $k_{\text{B}}T/h\nu$ is constrained to temperature range in which $\bar{n}\leq 1$, which is associated with the validity regime of the solution in Eq.~\eqref{Fifel}. It is important to emphasize the role played by the data-bus in reducing the decoherence effects in our system. It can be seen that for a given range of $k_{\text{B}}T/h\nu$, we can use the size of the data-bus as a strategic tool to enhance the performance of the system against the effects of a thermal environment. Second, the Fig.~\ref{Fig2}~(bottom) presents an analysis of the fidelity $\bar{\Fcal}$, as a function of the dimensionless parameter $\gamma/\lambda$ and $\kappa$, for different choices to the thermal reservoir temperature through the parameter $k_{\text{B}}T/h\nu$. As a complement to previous results, the Fig.~\ref{Fig2}~(bottom) suggests that we can not increase the data-bus indefinitely in order to get an optimal performance against decoherence. It establishes an optimal relationship between the data-bus size and the thermal reservoir parameters $(\gamma/\lambda,k_{\text{B}}T/h\nu)$, in which we could perform hardware engineering in order to minimize undesired thermal effects on our transistors. Such a result can be seen most clearly through the maximum point on the line separating the regions of density above and below 0.9 in Fig.~\ref{Fig2}~(bottom) for the parameters $k_{\text{B}}T/h\nu = 2\cdot10^{-1}$, $k_{\text{B}}T/h\nu = 5\cdot10^{-1}$ and $k_{\text{B}}T/h\nu = 1\cdot10^{0}$. In addition, from the Eqs.~\eqref{Fifel} and~\eqref{Fifelbar}, it is noted that the probability of success is independent on the phase-shift gate $\phi$ that will be implemented in our quantum device. Therefore, the results present in Fig.~\ref{Fig2} are valid for any $\phi$.

In order to give an experimental notion of how useful can be our transistor against decoherence effect, let us give a realistic example. Firstly, it is important to mention, in the rotating wave and Markov approximations, that the relevant coupling between each oscillator of the transistor with the thermal bath happens when the frequency $\nu$ is around to $\omega$~\cite{Scully,Walls}, where the characteristic value of $\omega$ in several system is of order of about $10$ GHz~\cite{circQED1,Cavity1}. With this approximate value of $\omega$ we can estimate the value of the temperature of the thermal bath from the quantity $k_{\text{B}}T/h\nu$ that appears on each graph in Fig.~\ref{Fig2}~(bottom). In fact, by using the experimental values of the constants~\cite{Mohr:16,Newell:18} $k_{\text{B}} =1.380 \cdot10^{-34}$~J~K$^{-1}$ and $h = 6.626\cdot10^{-23}$~J~s, for the case in which we have $k_{\text{B}}T/h\nu = 5\cdot10^{-1}$, for example, the temperature obtained will be $T \approx 0.24$~K. Thus, by considering the graph in Fig.~\ref{Fig2}~(bottom), one can conclude that, for the reservoir in which $\gamma/\lambda \leq 0.1$ and $T \leq 0.24$~K, the quantum transistor will work with high fidelity if we design a data-bus with approximately $10$ oscillators.

\section*{Discussion}

In this paper, we present a quantum transistor model based on quantum oscillators networks. We believe that it can be a useful device for the
quantum information processing with optical devices implemented
experimentally in both cavity-QED and circuit-QED, for example. Our model
explores the frequency detuning between the data-bus oscillators (the gate)
with the source and drain oscillators so that the data-bus allow us to create a ``potential barrier" to
block or transfer the quantum information from the source to the drain. In
this sense, the gate oscillators can be seen as an \textit{optical quantum switch}
for quantum information currents. In addition to blocking or transferring
quantum information (when the barrier is removed), the transistor proposed
here can be used to apply individual quantum gates when the
oscillator frequency, the coupling strength between the oscillators and the
number of resonant data-bus oscillators is properly adjusted. When
considering the inevitable coupling of the system with a thermal reservoir,
the performance of the transistor is dictated by the parameters of the
environment, namely, the bath temperature $T$ and the emission
rate $\gamma$. As expected, the system is strongly affected as the temperature $T$
increases. However, we can maintain the high fidelity transfers
(as well as the implementation of the phase-shift gate) for cavities with
low $\gamma$ emission rate or high quality factor. In particular, we have shown that the size of the data-bus ($\kappa$) can be used as a parameter to control the decoherence effects of the system. In the cases we consider here, we find graphically the existence of an optimal non-trivial criterion for the parameter $\kappa$, which depends on both the temperature and the spontaneous emission rate. The knowledge of this criterion allows us to design specific quantum devices where we can enhance the transfer/blocking fidelity against the effects of the thermal bath, in which the temperature and spontaneous emission rate are known. Obviously, the adaptability of our device depends heavily on our experimental ability to turn data-bus quantum oscillators on or off through the atom-field dispersive interaction. Provided that it can be done without too much difficulty, our model can be perfectly adapted to a wide variety of situations imposed by the thermal bath.

Since in our model we are interested in the weak coupling regime between the oscillators (i.e., $\sqrt{N}\lambda\ll \omega$), the rotating wave approximation can be performed. In the hypothesis that it is possible to implement the strong coupling regime between the oscillators, we know from literature \cite{Mickel:07-1,Mickel:07-2,Mickel:05,Mickel:04-1,Mickel:04-2} that there will be cross-dissipation channels that increase the fidelity of some particular state to be transferred or even eliminate the decoherence effect (dark-states). In view of this result, a natural extension of our work to the strong coupling regime should reveal us some additional gains in fidelity to some initial states and loss to others. As for the entanglement between the source-drain oscillators, what we expect, based on the references\cite{Mickel:07-1,Mickel:07-2,Mickel:05,Mickel:04-1,Mickel:04-2}, we knows that in a state recurrence time and/or state transfer time the entanglement degree goes to zero, because the states factorize from each other, and it is maximum in half this time, when we have a state entangled with the all data-bus oscillators. The fact that we have a reduction in the short time scale with the increase in the number of resonant data bus oscillators will only tell us that the degree of entanglement reaches its maximum value faster and not that the entanglement increases. This is because the topology of the network, which we propose in this article, is a sum of several transmission lines connected only by the extreme oscillators (each line comprises 3 oscillators). An understanding of the classical point of view can be made here: by increasing the number of these transmission lines, we reduce the fraction of the state to be transmitted between each line, reducing the short time scale and, according to the network adjustment, we can reduce the transfer time and consequently decrease the harmful effects of a thermal reservoir whose time scale remains unchanged.

Throughout this paper, we have studied a device that can be applied to
short-range communication, once we are interested in controlling quantum information within quantum devices. However, it is reasonable to believe that our
model could be extended to provide long-range communication, where it would
require a growth in the number of quantum oscillators or a change in data bus topology. We believe that our proposal opens perspectives for the
development of other schemes of optical quantum transistors, or more complex
optical devices derived from it. In addition, other approaches to the
development of new quantum transistors can be considered from the quantum
transistor models mentioned here. For example, the adiabatic quantum
transistor model \cite{Bacon:17} uses slow evolutions to accomplish the task
of transferring quantum information. In this sense, we can use adiabaticity
shortcuts \cite{Demirplak:03,Demirplak:05,Berry:09} to speed up this task,
where we could provide a \textit{superadiabatic} quantum transistor. Since
such STA method can be implemented in an arbitrary finite time \cite{Santos:15,Coulamy:16}, the use of advanced methods of STA \cite{Chen:18-3,Santos:18-b,Hu:18,Chen:10,XiaPRA:14,Garejev:14,Lu:14,Baksic:16,Huang:17,Chen:16-2}
to develop such quantum devices could be appreciated for superadiabatic
quantum computing \cite{Santos:15,Santos:16}. In addition, since this extended
model could be efficiently implemented using different physical systems \cite{Hennessy:07,Li:02,Zhang:17,Irish:08,Xiao:08,Eisert:15,Kaufman:16,Tomita:17}, a theoretical and experimental studies will be considered in future researches.

\section*{Methods}

For our purposes, we will consider a data bus consisting of a network of $N$ non-interacting oscillators, which, however, each one is coupled with the source and drain oscillator with a real coupling strength $\{\lambda_{sm}\}=\{\lambda_{dm}\}=\lambda$. In addition, let's consider that the frequencies of $\kappa$ data-bus oscillators are in resonance with the frequencies of the source and drain, $\omega_s=\omega_d=\omega$, while the others data-bus oscillators have dissonant frequencies $\tilde{\omega}=\omega+\Delta$. Under these conditions, we can obtain analytical expressions that are written in terms of the eigenvalues and eigenvectors of the matrix $\mathcal{H}$, defined by
\begin{equation}
\mathcal{H}=\left( 
\begin{array}{ccc}
\omega & \mathbf{\Lambda } & 0 \\ 
\mathbf{\Lambda }^{\top } & \mathcal{H}^{\text{DB}} & \mathbf{\Lambda }^{\top } \\ 
0 & \mathbf{\Lambda } & \omega
\end{array}\right) ,  \label{A1}
\end{equation}%
where $\mathbf{\Lambda}$ is a $1\times N$ matrix, whose elements are $\Lambda_{m}=\lambda$, $\mathbf{\Lambda}^{\top}$ is the transposed matrix of $\mathbf{\Lambda}$, whereas the square $N\times N$ matrix $\mathcal{H}^{\text{DB}}$ is a diagonal matrix whose elements are defined by $\mathcal{H}^{\text{DB}}_{mn}=\omega\delta_{mn}$, if $1\leq m\leq \kappa$ and $\mathcal{H}^{\text{DB}}_{mn}=\tilde{\omega}\delta _{mn}$, if $\kappa <m\leq N$. According to this matrix, the Hamiltonian
\begin{equation}
H=\hbar \left[ \omega \left( \sum_{\ell =s,d}a_{\ell }^{\dag }a_{\ell
}+\sum_{m=1}^{\kappa }a_{m}^{\dag }a_{m}\right) +\tilde{\omega}%
\sum_{m=\kappa +1}^{N}a_{m}^{\dag }a_{m}+\lambda \sum_{\ell
	=s,d}\sum_{m=1}^{N}\left( a_{\ell }^{\dag }a_{m}+a_{\ell }a_{m}^{\dag
}\right) \right]  \label{H1}
\end{equation}%
can be put into the matrix form as
\begin{equation}
H=\left( 
\begin{array}{ccc}
a_{s}^{\dag } & \left\{ a_{m}^{\dag }\right\} & a_{d}^{\dag }%
\end{array}%
\right) \left( 
\begin{array}{ccc}
\omega & \mathbf{\Lambda } & 0 \\ 
\mathbf{\Lambda }^{\top } & \mathcal{H}_{DB} & \mathbf{\Lambda }^{\top } \\ 
0 & \mathbf{\Lambda } & \omega%
\end{array}\right) \left( 
\begin{array}{c}
a_{s} \\ 
\left\{ a_{m}\right\} \\ 
a_{d}%
\end{array}%
\right) \text{.}  \label{H2}
\end{equation}

Considering that $j$ and $j'$ vary from $0$ to $N + 1$, the eigenvalues and orthonormal eigenvectors of $\mathcal{H}$ can be labeled as follows:
\begin{itemize}
	\item  Regardless of the value of $\kappa$ we will always have an eigenvalue $\Omega_{0}=\omega$, whose eigenvector $\vartheta_{0}$ has the components $C_{j0}=1/\sqrt{2}$, if $j=0$; $C_{j0}=-1/\sqrt{2}$, if $j=N+1$; and $C_{j0}=0$ for any other value of $j$.
	
	\item When $2\leq\kappa\leq N$ we find $\kappa -1$ identical eigenvalues, which can be labeled by $j'$ as follows: For $1\leq j'\leq\kappa-1$, we obtain the eigenvalue $\Omega_{j'}=\omega$, whose eigenvector $\vartheta_{j'}$ has the components $C_{jj'}=1/\sqrt{\left(j +1\right) j }$, if $1\leq j\leq
	j'$; $C_{jj'}=-(j-1)/\sqrt{j \left(j-1\right)}$, if $j=j' +1$; and $C_{jj'}=0$ for any other value of $j$.
		
	\item When $0\leq\kappa\leq N-2$ we find $N-\kappa-1$ equal eigenvalues, which will be labeled by $j'$ as follows: For $\kappa\leq j'\leq N-2$, we have the eigenvalue $\Omega_{j'}=	\tilde{\omega}$, whose eigenvector $\vartheta_{j'}$ has the components $C_{jj'}=1/\sqrt{(j-\kappa+1)(j-\kappa)}$, if $\kappa +1\leq j\leq j' +1$; $C_{jj'}=-(j-\kappa-1)/\sqrt{(j-\kappa)(j-\kappa-1)}$, if $j=j'+2$; and $C_{jj'}=0$ for any other value of $j$.
	
	\item Setting the parameters $\Phi=\Delta^{2}+6N\lambda^{2}$, $\eta=\Delta \left[\Delta^{2}+9\left(N-3\kappa\right)\lambda^{2}\right]$, and $\theta =\frac{1}{3}\arctan\left(\sqrt{\frac{\Phi^{3}}{\eta ^{2}}-1}\right)$,	we can write the last three eigenvalues in compact form as
		\begin{eqnarray}
		\Omega_{j'} =\omega +\frac{\Delta }{3}-(1-3\delta_{j',N})\frac{\sqrt{\Phi }}{3}\cos\theta+(1-\delta_{j',N}-2\delta_{j',N-1})\sqrt{\frac{\Phi}{3}}\sin\theta,
		\end{eqnarray}
	where $j'=N-1,N,N+1$. The eigenvector $\vartheta_{j'}$ associated with each of these eigenvalues has coefficients defined by: $C_{jj'}=\mathcal{N}_{j'}$, if $j=0$ or $j=N+1$; $C_{jj'}=2\lambda\mathcal{N}_{j'}/(\Omega_{j'}-\omega)$, if $1\leq j\leq\kappa$; and $C_{jj'}=2\lambda\mathcal{N}_{j'}/(\Omega{j'}-\tilde{\omega})$, if $\kappa<j\leq N$, where we define the normalization coefficient by
		\begin{equation}
		\mathcal{N}_{j'}=\left[2+\kappa\left(\frac{2\lambda}{\Omega_{j'}-\omega }\right) ^{2}+(N-\kappa)\left(\frac{2\lambda }{\Omega_{j'}-\tilde{\omega}}\right)^{2}\right]^{-1/2}\text{.}
	\end{equation}
\end{itemize}

Once the matrix $\mathbf{C}$ is obtained, whose columns are the eigenvectors of $\mathcal{H}$, we can diagonalize the Hamiltonian so that $\mathbf{C}^{-1}\cdot \mathcal{H}\cdot \mathbf{C}=\mathcal{H}_{D}$, where the elements of the diagonal matrix $\mathcal{H}_D$ are the eigenvalues defined above. The new $A$ operators, which follow the same canonical commutation rules as the original operators $a$, are defined by
\begin{equation}
A_{j'}=\sum_{j}C_{j'j}^{-1}a_{j},
\end{equation}
remembering that $\mathbf{C}^{-1}=\mathbf{C}^{\top }$ and that we define $a_{0}=a_{s}$ and $a_{N+1}=a_{d}$.

In the situation where we have a thermal reservoir coupled to
each of the oscillators of our system, according to Ref.\cite{Mickel:07-1}, we can write the temporal evolution through the elements of a matrix $\mathbf{\Theta}(t)$, defined by:
\begin{equation}
\Theta_{j'j}\left( t\right) =e^{-\gamma t/2}\sum_{l=0}^{N+1}C_{j'l}e^{-i\Omega_{l}t}C_{lj}^{-1}.
\end{equation}
The diffusion of the system occurs due to the presence of the matrix $\mathbf{J}(t)$, which for situations in which the reservoirs are identical, that is, they have the same spontaneous decay rate $\gamma_j=\gamma$ and the same average number of thermal photons $\bar{n}_j=\bar{n}$, can be written as
\begin{equation}
	J_{j'j}\left(t\right)=2\bar{n}\left(1-e^{-\mathbf{\gamma }t}\right)\delta
	_{j'j}\text{.}
\end{equation}

If we consider the initial state of the system given by $ \vert\psi
_{\text{sgd}}\left(0\right) \rangle=\left(b_{0}\left\vert
0_{s}\right\rangle +b_{1}\left\vert 1_{s}\right\rangle \right) \otimes
 \vert \{0_{g}\},0_{d} \rangle $, we will verify that the time evolution of the density operator becomes\cite{Mickel:07-1}
\begin{eqnarray}
\rho _{\text{sgd}}(t) &=&\sum_{m,n=0}^{1}b_{m}^{\ast
}b_{n}\sum_{l=0}^{n}\prod\limits_{\ell =0}^{N+1}\left\{ \sum_{r_{\ell
},s_{\ell },k_{\ell }=0}^{\infty }\sum_{S_{\ell }=0}^{\min \left\{ r_{\ell
},s_{\ell }\right\} +k_{\ell }}\left( -1\right) ^{k_{\ell }}\frac{\left[
\Theta _{\ell 0}\left( t\right) \right] ^{r_{\ell }+k_{\ell }-S_{\ell }}%
\left[ \Theta _{\ell 0}^{\ast }\left( t\right) \right] ^{s_{\ell }+k_{\ell
	}-S_{\ell }}\left[ \bar{n}\left( 1-e^{-\mathbf{\gamma }t}\right) \right]
^{S_{\ell }}}{\left( r_{\ell }+k_{\ell }-S_{\ell }\right) !\left( s_{\ell
}+k_{\ell }-S_{\ell }\right) !S_{\ell }!}\right.   \notag \\
&&\times \left. \frac{\left( r_{\ell }+k_{\ell }\right) !\left( s_{\ell
	}+k_{\ell }\right) !}{k_{\ell }!\sqrt{r_{\ell }!s_{\ell }!}}\left\vert
r_{\ell }\right\rangle \left\langle s_{\ell }\right\vert \right\} \delta %
\left[ m-n+\sum_{\ell =0}^{N+1}\left( r_{\ell }-s_{\ell }\right) \right]
\delta \left[ l-\sum_{\ell =0}^{N+1}\left( r_{\ell }+k_{\ell }-S_{\ell
}\right) \right] \text{,}
\end{eqnarray}
where we are considering the fact that $\delta[x]=1$ if $x=0$ and $\delta[x]=0$ if $x\neq 0$.


\section*{Acknowledgements}

We acknowledge financial support from the Brazilian agencies CNPq and the Brazilian National Institute of Science and Technology for Quantum Information (INCT-IQ). Specially, ACS would like to thank Dr. Marcelo Silva Sarandy (IF/UFF) and MAP would like to thank GILY for useful discussions. Both authors would like to express sincere thanks to Gislaine C. Batistela (UNESP), Belita Koiller (IF/UFRJ) and Kita Macario (IF/UFF) for being so solicitous when requested.

\section*{Author contributions statement}

Both authors Alan C. Santos and M. A. de Ponte conceived the idea, developed the calculations, discussed the results and wrote the manuscript.

\section*{Competing interests statement }

There are no conflicts of interest, of any nature, for both authors of this article.

\end{document}